# PHYSICS TEACHING AND EDUCATIONAL INTERDISCIPLINARITY WITH A.V. USOVA REVISITED

## O. Yavoruk

*Independent Scholar (RUSSIAN FEDERATION)*

## Abstract

The paper deals with pedagogical heritage of A.V. Usova (1921-2014) that yields insights into problems of interdisciplinary relationship in the school science education system. Her fundamental works were used as teaching guides in the preparation of physics teachers of all pedagogical universities of the Soviet Union and Russia. In this study, professor Usova is not an interdisciplinary researcher; she is a researcher of interdisciplinarity. Here is provided a short and consistent description of her views on teaching physics as a foundation of educational interdisciplinarity. Usova elaborated educational functions, classifications, organizational levels, ways of instantiating, content-related foundations of interdisciplinary relations, recommendations for practical applying interdisciplinary relations. Besides, Usova suggested a unified approach to the study of knowledge and skills through the generalized plans. And we would like not simply clarify, retell or restate the very powerful didactic discoveries of Usova, but consider opportunities of teaching across boundaries for actual educational challenges. The future of the field of didactics, which deals with the study of interdisciplinarity, is very interesting and attractive. This problem will always excite scientists, educators and teachers, and the futurity of science education depends on the ways it is solved.

## 1 INTRODUCTION

2021 marks the 100th anniversary of the birth of an outstanding Soviet and Russian scientist and pedagogue Antonina Vasilyevna Usova. Any round-number anniversary encourages deep rethinking the previous experiences, current problems and views on the future. Understanding of happened events and accomplished transformations comes after some time, and the feeling of being late always does not leave us (everything is done with delay). We want to understand what is completely out of date, what has significantly changed, and what is still essential and relevant.

A.V. Usova is the most important educational theorist in physics teaching of Russian education. She made crucial contributions to the different fields of pedagogical science: general and specific issues of physics teaching at schools and universities, concept learning in physics classes, increasing of interest in the study of physics, organization of students' self-learning, formation of general learning skills, teaching students to solving physics problems, methodology of pedagogical experiment (elemental and operational analysis), issues of polytechnic education, theory and practice of developing education, educational interdisciplinarity.

A.V. Usova called herself a teacher, a scientist, and a didactician (Didactics is "the activities of educating or instructing; activities that impart knowledge or skill" [1] and "the art or science of teaching" [2]). In the history of pedagogical sciences, she remained as the author of many fundamental didactic ideas.

Attempts to comprehend Usova's contribution into pedagogical sciences were undertaken by many researchers [3], [4], [5]. Her pedagogical views were discussed at scientific conferences [6].

A.V. Usova published her books and articles in Russian only, and this fact complicates our review, since a simple transfer of pedagogical traditions, terminology and approaches of Soviet and Russian education is impossible. Some educational terms cannot be rendered with complete accuracy from one linguistic system to another. We detect certain translation difficulties with terms that concern practical physics teaching. They ("metodika", "medodicheskiye", "metodika prepodavaniya fiziki", "obobshchyonnye plany") do not have accurate English equivalents. Some of her books have been translated into Spanish [7] and Bulgarian [8].

Different languages and different cultures often don't have comparable terms. There are small but important differences in using these terms. Teachers and educators still argue about these definitions. We are going to use the term "interdisciplinary relations" since it is closest to the ideas of Usova. Of course, relations concern different kinds of connecting, integrating and unifying several scientific disciplines.

Word-by-word translation will not achieve the desired goals for many reasons. Nowadays, a lot of things may seem naive and outdated. Eventually, our world is open, and the interested researcher can personally read the primary sources anyway. My retellings of Usova's contribution to the field of educational interdisciplinarity are based not only on texts, but also on personal conversations with Usova.

## 2 INTERDISCIPLINARY TEACHING BY USOVA

Usova's achievements in the study of interdisciplinarity can be used in the process of teaching different scientific disciplines [9]. Her brochure "Interdisciplinary relationship in school science teaching" was reprinted several times [10]. The book on the methods of teaching physics, also published many times, necessarily included a chapter on interdisciplinary relations [11].

A.V.Usova regularly published thematic collections of articles [12], [13], [14] and organized topical conferences [15]. The interdisciplinary relationship was a crucial factor in Usova's original Conception of Natural Science Education [16], [17], [18]. And we should note the interest of A.V. Usova to the history of physics [19].

The problems of interdisciplinary didactics were regularly discussed on the conference series "Methodology of the scientific concepts formation among school and university students". Mastermind of these conferences was A.V. Usova. She organized them annually (Chelyabinsk, South Urals, Soviet Union and Russia, 1971-2013) [20].

Year after year Usova developed and improved her ideas, clarifying formulations, perfecting statements, refining phrases, taking into account emerging circumstances. Her apprentices have developed various aspects of interdisciplinary teaching (integrated disciplines, multidisciplinary classes, complex seminars, integrative lessons, interdisciplinary lectures, practical works, labs, exercises, etc.) with detailed descriptions, step-by-step instructions and guides [21], [22], [23], [24], [25], [26], [27], [28], [29], [30], [31].

Interdisciplinary activity is often accused of dilettantism and "shoddy standards" [32]. These statements do not sound particularly surprising, and sometimes we see the resistance to interdisciplinary ideas in the scholar discussions. Academic disciplines are self-contained. Without trying interrelations they will always be receding from each other. That lets us take for granted that what we consider to be our way of disciplinary teaching is actually reliable. Interdisciplinarity is a particular way of teaching that is still far from clear. We understand disciplinary knowledge fairly well, what can we lose or gain due to removing interdisciplinarity? It is hard but we cannot say that it is definitely impossible.

Usova understood clearly the importance of school physics teaching and was a strong supporter of the disciplinary structure of natural science teaching. She came up with a wonderful definition of what interdisciplinary relations are. In her educational conception interdisciplinary relations was a didactic condition for disciplinary teaching, the quality of students' knowledge improving, development of thinking, worldview forming, technical preparing, and professional orientation. It is a remedy to the harmful effects of educational disciplinary specialization.

It is easy to understand what Usova had in mind. Physics and some disciplines have a longstanding relationship. Their relation may be characterized as a sort of friendly rivalry. These are astronomy, biology, chemistry, mathematics, history, philosophy, psychology, sociology, etc. They have different aims, methods and results, but their subjects often overlap. And it is ultimately interesting to view the same object from different points of view, comparing the results. Sometimes it's funny and weird, but results are always enlightening. So how can the system ever adequately get going? The answer is only due actuating relations between separate components (disciplines).

Moreover, she encourages to the earlier study of school physics, and this propaedeutic (introductory) physics course has many passionate proponents [33]. Physics can serve as a pillar in clarifying interdisciplinary relationships by virtue of its fundamentality. Physics is viewed as the core and origin of natural science education.

In a system of education in which there were no regular relations, there would be no good knowledge and confidence in its usefulness. But what do interrelations really mean, and where does it come from? From an epistemological point of view, interdisciplinary relations are a reflection of relations between sciences. Science has a disciplinary structure, but complex problems solving often requires understanding of different disciplines. Besides we often deal with complex subjects that should be examined by multiple scientific fields.

However, the representation of the same concepts (vector, velocity, etc.) in physics and mathematics textbooks can be radically different. Reasoning about many other concepts and laws in physics do not match up with the same in mathematics, chemistry, biology, etc.

Usova considered educational functions, classifications, ways of instantiating, content-related (contentual) foundations of interdisciplinary relations [34]. Here are some Usova's key recommendations for practical applying interdisciplinary relations.

✦ *The directions of teachers' interdisciplinary activity (according A.V. Usova)*
  1. The temporal coordination of scientific disciplines: teaching every discipline should be based on the previous ones.
  2. The continuity support in teaching general concepts, laws, and theories.
  3. The unity in the knowledge interpretation (for learning concepts, laws, and theories).
  4. Unified approaches for developing an interdisciplinary skills.
  5. The use of material from other disciplines in physics classes.
  6. The demonstration of the unity for research methods.
  7. Solving physics problems with a complex application of scientific knowledge.
  8. Elimination of duplicating the same issues in different disciplines.
  9. Designing and delivering complex interdisciplinary classes.

According to A.V. Usova it is necessary to fulfill a certain set of regular interdisciplinary activity by all participants of the educational process (on all organizational levels). Each hierarchical level of interdisciplinarity manifests a distinct role and that they coexist in a dynamic unity.

✦ *Organizational levels of the interdisciplinary relationships fulfillment*
  1. Federal, regional, school commissioners developing curricula.
  2. Syllabus commissioners and educators who develop ways and tools of implementing interdisciplinary relations in teaching specific disciplines.
  3. Authors of textbooks for students and guides for teachers.
  4. School administrators and interdisciplinary commissioners.
  5. School teachers.

Besides, Usova suggests a unified approach to the study of knowledge and skills on the basis of generalized plans.

✦ *Physics Phenomena*
  1. External features of the phenomenon.
  2. Conditions of the phenomenon.
  3. The essence of the phenomenon, its mechanism.
  4. Relation of the phenomenon with other phenomena.
  5. Quantitative characteristics of the phenomenon.
  6. Application of the phenomenon.
  7. Prevention of harmful effects of the phenomenon.

✦ *Physics Quantities*
  1. What phenomenon or property is characterized by this quantity?
  2. Definition of the physics quantity.
  3. Specific characteristics of this quantity (scalar or vector, basic or derivative)
  4. Definitional formulas for this quantity.
  5. Units of this quantity according to SI (International System of Units).
  6. Methods of measuring.

- *Laws of Physics*
    1. What concepts are related to the law of physics?
    2. Formulations of the law.
    3. The mathematical expressions of the law.
    4. Limitations of the law.
    5. Who discovered the law? When and how?
    6. Experimental evidence of the law.
    7. Examples of natural phenomena, where the law acts.
    8. Practical application of the law.
- *Theories of Physics*
    1. Scientific facts on which the theory is based.
    2. Key concepts of the theory.
    3. The basic assertions, postulates, laws, principles of the theory.
    4. Mathematical formalism of the theory, fundamental equations.
    5. Results, findings, and predictions of the theory.
    6. Limitations of the theory.
- *Plan of Observations*
    1. Information about the observer.
    2. Date/time of the observation, its duration.
    3. Purpose of the observation (it is formulated by a teacher or chosen by the students).
    4. Equipment of the observation (it may be selected by students).
    5. Conditions of the observation.
    6. Description of the observational procedure.
    7. Description of the observed object.
    8. Results of the observation.
    9. Analysis of results and conclusions from them.

Describing the generalized plans, Usova necessarily emphasized their interdisciplinary status. She has posited that manipulating generalized plans in such a way as to create learning activity will inevitably create some sort of positive feedback.

## 3   REASONING AND RETHINKING ON ACTUAL ASPECTS OF EDUCATIONAL INTERDISCIPLINARITY

Usova's views are easily applicable in contemporary interdisciplinary teaching. Her reasoning is understandable for modern researchers of interdisciplinarity.

The category "relationship" is one of the most used categories in science at the end of the 20th and beginning of the 21st century. This category is widely used in pedagogical sciences too: numerous natural and artificial relations permeate educational process, showing us their amazing internal integrity, consistency and interdependence. Here is discussed a special and very clearly distinguished group of relations, which is called "interdisciplinary relations" in psychological and didactic literature. Didactic interdisciplinarity as a condition of teaching is the Usova's way of formalizing this idea. Usova's opinion about this kind of interdisciplinarity (and how it may be used) is fairly compelling.

It is clear that spontaneous relations are often wrong, and the aimless interdisciplinary teaching activity turns out to be poorly effective. Ignoring tangible interdisciplinary relations can produce deleterious outcomes: a very important and fundamental part of human knowledge is outside the school curriculum. As a result, the following negative phenomena are observed: students feel insecure about academic knowledge, they do not know how to apply in practice, they do not distinguish between scientific knowledge and unscientific one, and the same concepts are interpreted differently in different disciplines.

Some teaching guides ignore issues related to the interdisciplinarity, or only declare it. The obvious disunity and disorganization explain some surprising and seemingly counterintuitive phenomena involving disciplinary teaching.

Disciplinary oriented teachers draw clearly defined boundaries where they are not in nature, forming general scientific skills (observation, experimentation) and knowledge (facts, concepts, laws, theories) in a contradictory way. Interdisciplinary relations are extra ingredient of disciplinary teaching. And teachers have argued about interdisciplinarity ever since the idea of disciplinary separation came up with planning the teaching ways. The existence of scientific disciplines, the separate fields of teaching (mathematics teaching, physics teaching, chemistry teaching, biology teaching) determines the emergence of the educational interdisciplinarity, the branch of interdisciplinary teaching or *Interdisciplinary Didactics.*

The next six aspects of the interdisciplinary didactics can be considered as a basis for discussions on teachers' meetings or a plan of self-education. This is the answer to the question: "What should we know about educational interdisciplinarity?"

### 3.1 Sources and factors of interdisciplinary didactics

These are the processes of crossing the boundaries and deepening relations between sciences, the emergence of scientific conceptions that are important for all branches of human activity. The most outstanding discoveries of our time occur at the intersection of sciences. Developing knowledge is formed into new scientific systems, stimulating us to search for its adequate application in teaching.

### 3.2 Psychological foundations of interdisciplinarity

A special issue is identifying the psychological foundations of interdisciplinary teaching. Keywords in the study of the psychological mechanisms of interdisciplinarity are the concepts of "thinking", "learning activity", "connection", "association", "integration", and "synthesis". Despite the results already achieved, building the psychological foundations of interdisciplinarity is not completed.

### 3.3 Interdisciplinary knowledge to learn

Usova recommends a universal approach to teaching knowledge widely applying generalized plans to the content-related foundations of physics, chemistry, biology, etc. The elements of science knowledge, identifying by Usova: scientific facts, concepts, laws, theories, etc. Their universality is the base for a unified approach to the study of different scientific disciplines [10]. Students can find out interdisciplinary knowledge in sources for additional reading (books, websites), popular scientific films, literature, teacher stories, and interdisciplinary projects.

### 3.4 Interdisciplinary skills to acquire

These are universal skills which are common to different scientific disciplines (kinds of students' activity concerning: textbook reading, observations and experimentations, systematizing and generalizing knowledge, etc.) [9]. Usova and her apprentices usually consider of a general series of general interdisciplinary skills: cognitive, practical, organizational, control, evaluative ones [22].

### 3.5 Organizational aspects of interdisciplinarity

The real question most teachers are interested in, however, is: how can I do it in my everyday teaching practice? A.V. Usova identifies where and how interdisciplinarity works, the different organizational levels of practical interdisciplinarity, and the teacher can be active only on one of them [10].

### 3.6 Prospects and opportunities for interdisciplinarity in the future

Interdisciplinarity issues ultimately result from a situation where we consider an object from the point of different disciplines views. These points of views can contradict to each other. We can never ignore this educational phenomenon.

Interdisciplinarity plays an important role in the development of students' thinking [35], as well as in the formation of scientific concepts [36]. This is a crucial clue to us when we talk about the future of interdisciplinarity.

Usova had her own opinion about the problem of educational integration: short generalizing integrative course should finalize the study of natural sciences at secondary school; integrative lessons are one of the ways of interdisciplinarity actuating. The ratio of concepts "Relation" and "Integration" is essential for Usova. According to Usova's version of the interdisciplinary didactics "Integration" is a special (particular) case of "Relation". In disputes on integration issues Usova always clearly distinguished the concepts "Integrative" and "Integrated".

The general prospects and opportunities of educational interdisciplinarity depend on the development and demands of science, technology and society.

## 4  CONCLUSIONS

Interdisciplinary issues of science teaching have their genesis in the science, have been evolving due to durable applying into educational processes, and remain developmentally open-ended.

The traditional and regular disunity of scientific disciplines is not something we can take for granted. When we talk about unity, we'll see that there is a well-defined scientific sense in which different people can clearly understand different approaches to nature, but the only way we can do that is by being on different lessons with different points of view, different teachers, and different schools.

The promising version of interdisciplinary teaching was presented us by Antonina Vasilyena Usova, a soviet and Russian pedagogue. Usova realized that interdisciplinary relations can be considered as a condition of the educational process. It does not stand as a separate feature of Usova's pedagogy but rather goes hand-in-hand with her views to physics teaching and concept learning.

In Usova's formulation we should teach fundamental scientific disciplines, improve the quality of students' knowledge and skills, develop thinking, form a worldview, deal with polytechnic education, and facilitate vocational guidance, provided interdisciplinary relationship. It is the job of educators to design educational structures so these conditions are met, thus ensuring that structures will be in interdisciplinary equilibrium and (maybe) some kind of acceptable unity. At the same time a centre of interdisciplinarity is physics.

The bottom line here is this. Looking to the past, we try to understand the pretty unpredictable future. What we really mean is the problem of educational interdisciplinarity does not have a final decision. But sooner or later, new researchers will appear with a much more subtle way of thinking about interdisciplinarity. What will they offer? How will they rate our experience? These questions remain open.